\documentclass[fleqn,10pt]{wlscirep}
\usepackage[utf8]{inputenc}
\usepackage[T1]{fontenc}
\title{Clustering patterns in efficiency and the coming-of-age of the cryptocurrency market}

\author[1]{Higor~Y.~D.~Sigaki}
\author[2,3]{Matja{\v z} Perc}
\author[1]{Haroldo~V.~Ribeiro}
\affil[1]{Departamento de F\'isica, Universidade Estadual de Maring\'a, Maring\'a, PR 87020-900, Brazil}
\affil[2]{Faculty of Natural Sciences and Mathematics, University of Maribor, Koro{\v s}ka cesta 160, SI-2000 Maribor, Slovenia}
\affil[3]{Complexity Science Hub Vienna, Josefst{\"a}dterstra{\ss}e 39, A-1080 Vienna, Austria}

\begin{abstract}
The efficient market hypothesis has far-reaching implications for financial trading and market stability. Whether or not cryptocurrencies are informationally efficient has therefore been the subject of intense recent investigation. Here, we use permutation entropy and statistical complexity over sliding time-windows of price log returns to quantify the dynamic efficiency of more than four hundred cryptocurrencies. We consider that a cryptocurrency is efficient within a time-window when these two complexity measures are statistically indistinguishable from their values obtained on randomly shuffled data. We find that 37\% of the cryptocurrencies in our study stay efficient over 80\% of the time, whereas 20\% are informationally efficient in less than 20\% of the time. Our results also show that the efficiency is not correlated with the market capitalization of the cryptocurrencies. A dynamic analysis of informational efficiency over time reveals clustering patterns in which different cryptocurrencies with similar temporal patterns form four clusters, and moreover, younger currencies in each group appear poised to follow the trend of their `elders'. The cryptocurrency market thus already shows notable adherence to the efficient market hypothesis, although data also reveals that the coming-of-age of digital currencies is in this regard still very much underway.
\end{abstract}
\begin{document}

\flushbottom
\maketitle

\thispagestyle{empty}

\section*{Introduction}

The efficient market hypothesis in financial economics asserts that asset prices fully reflect all available information~\cite{malkiel1970efficient}. A market is thus said to be informationally efficient if asset prices are always fully up-to-date, such that any new information about a particular firm is certain and immediately priced into its stock. This of course has important implications for financial trading in that it is impossible to beat an informationally efficient market with expert stock selection or arbitrage. Moreover, the efficient market hypothesis effectively prevents economic bubbles, which are amongst the main culprits behind stock crashes and market instability~\cite{preis2011bubble, sornette2017stock}. On the other hand, critics of the efficient market hypothesis claim that it is precisely the (false) belief in rational markets that is to blame for the '07-'08 financial crisis, as well as for many others unwanted developments in the World economy~\cite{fox2009myth}.

Whether the efficient market hypothesis is the Holy Grail of financial economics, or whether it is just a simplification of reality that does not always hold but is still valid for most investment purposes -- fact is that it is an immensely important concept that, even half a century since its inception, still critically shapes today's economic thought. Not surprisingly, recent research has been hard at work in probing whether the newly introduced digital currencies stand the test of the efficient market hypothesis~\cite{urquhart2016inefficiency, bariviera2017some, zhang2018inefficiency, bariviera2017inefficiency, nadarajah2017inefficiency, tiwari2018informational, bariviera2018analysis, alvarez2018long} (as well as other questions~\cite{elbahrawy2017evolutionary,javarone2018from,ermann2018google,li2018sentiment,alessandretti2018anticipating}). One of the earliest publications in this direction by Urquhart~\cite{urquhart2016inefficiency} reported that Bitcoin returns are significantly inefficient over the whole studied sample, but when split into two subsample periods, tests indicated that efficiency is there in the second period. Subsequently, Bariviera~\cite{bariviera2017inefficiency} arrived at similar conclusions using a dynamic approach, which revealed that the Bitcoin market seems to be more informationally efficient from 2014 onwards. Informational efficiency of the Bitcoin market was also confirmed by Nadarajah and Chu~\cite{nadarajah2017inefficiency}, and by Tiwari \textit{et al.}~\cite{tiwari2018informational}.

The growing popularity of cryptocurrencies, despite volatile prices, indicates that decentralized control through the blockchain technology, along with secure financial transactions due to strong cryptography, are highly valued among customers worldwide. A better overall understanding of the digital currency market is therefore much needed. To that effect, we here conduct a large-scale efficiency analysis of this market, analyzing the daily price time series of 437 cryptocurrencies. Methodologically, we rely on the permutation entropy~\cite{BandtPompe2002} and statistical complexity~\cite{LopezManciniCalbet1995,RossoLarrondoMartinPlastinoFuentes2007} over sliding time-windows of price log returns. Past research has shown that such a physics-inspired approach works very well on economic data~\cite{mantegna1999introduction}, in particular for quantifying the stock market inefficiency~\cite{zunino2007inefficiency,zunino2008multifractal, zunino2009forbidden, zunino2010complexity,zunino2012efficiency}. Once determining the permutation entropy and statistical complexity, we consider that a cryptocurrency is informationally efficient within a time-window when both measures are within the 95\% random confidence interval. We obtain the latter by shuffling the time series in each window and calculating the permutation entropy and statistical complexity for several independent realizations.

As we shall show in what follows, our research reveals that 37\% of the 437 cryptocurrencies are informationally efficient in over 80\% of the time, while 20\% stay efficient in less than 20\% of the time. We also find that efficiency is not correlated with the market capitalization of the cryptocurrencies. Moreover, a dynamic analysis of informational efficiency over time reveals clustering patterns in which different cryptocurrencies are grouped together due to their similar temporal profiles. For the clustering analysis, we rely on dynamic time warping~\cite{sakoe1978dynamic}, which has the advantage that distances between time series of different lengths and scale of values can still be determined. Based on this analysis, we find four groups of cryptocurrencies: \textit{i)} those that begin at a higher efficiency level but evolve to a lower one (12\% of market); \textit{ii)} those that improve the efficiency over time (19\% of market); \textit{iii)} those displaying a nearly constant efficiency level (43\% of market); and those that start at a higher efficiency level, decrease to a lower level, and then increase their efficiency level (26\% of market). This clustering analysis also indicates that younger currencies in each group appear poised to follow the trends of their `elders'. Overall, we thus find that large parts of the cryptocurrency market satisfy the efficient market hypothesis most of the time, but also that the coming-of-age of digital currencies is in this respect still in progress.

\section*{Results}

\begin{figure*}[!ht]
\centering
\includegraphics[width=0.98\columnwidth, keepaspectratio]{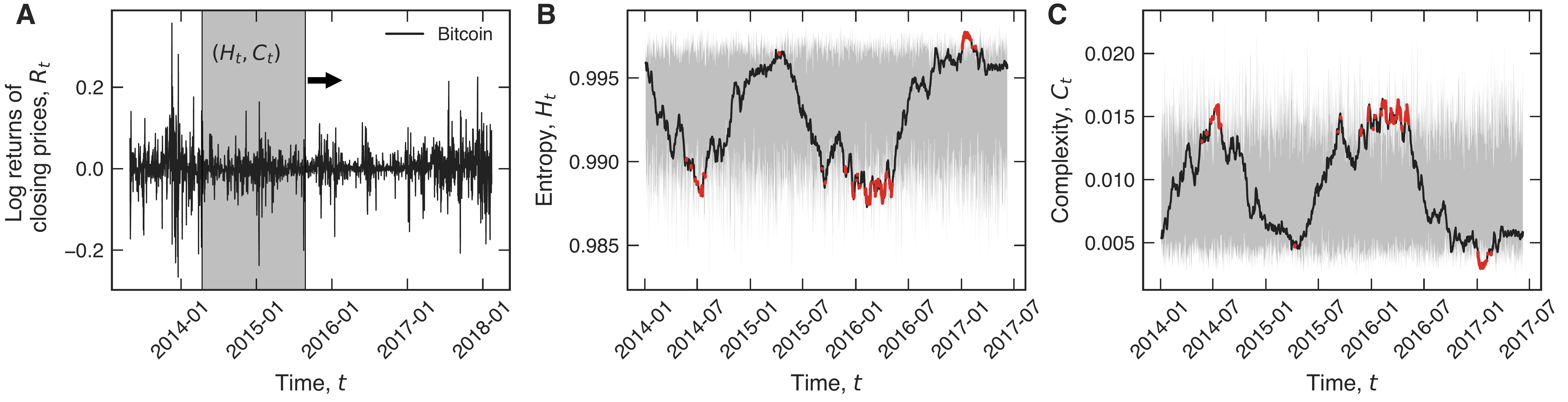}
\caption{Quantifying the informational efficiency of crypto-assets with permutation entropy and statistical complexity. (A) Time series of log returns $R_t$ of the closing prices for Bitcoin (BTC) since April 28, 2013. The shaded area shows a 500-day (about two years of economic activity) sliding window, which moves forward by 1 data point. The black curves in panels (B) and (C) show the time evolution of the permutation entropy $H_t$ and statistical complexity $C_t$ estimated within the sliding window, respectively. The embedding dimension used was $d=4$ (Methods Section for details). The shaded areas represent the 95\% random confidence intervals obtained by shuffling the data in each window and calculating the values of the entropy and complexity over several independent realizations. The red segments indicate the values of $H_t$ and $C_t$ that are outside the random confidence band. We define the overall informational efficiency $E$ as the fraction of time (days) that both complexity measures stay within the 95\% random confidence band. The estimated efficiency for Bitcoin is $E\approx0.85$ in the period under analysis.}
\label{fig:1}
\end{figure*}

Our results are based on data comprising the daily closing prices and market capitalization of 437 cryptocurrencies, including the most popular and important crypto-assets in circulation today (Methods Section for details). We begin by presenting our methodology on the Bitcoin (BTC) time series, arguably still the most well-known and popular cryptocurrency. As shown in Fig.~\ref{fig:1}A, the log return $R_t$ of the closing price time series is sampled with a sliding window (shaded gray) comprising 500 data points, which corresponds to approximately two years of economic activity. The sliding window moves ahead one day at a time, and for each, we determine the permutation entropy~\cite{BandtPompe2002} $H_t$ and the statistical complexity~\cite{LopezManciniCalbet1995,RossoLarrondoMartinPlastinoFuentes2007} $C_t$, as shown in Figs.~\ref{fig:1}B and C, respectively (Methods Section for details). These two complexity measures are estimated from the local ordering patterns among consecutive values of $R_t$. The permutation entropy measures the degree of randomness in the occurrence of these patterns, ranging from $H\approx 0$ for a completely regular series to $H\approx 1$ for a completely random series. The statistical complexity, in turn, quantifies the structural complexity in this ordering dynamics: $C\approx 0$ in both extremes of order and disorder, whereas $C>0$ when the ordinal patterns occur in a more complex fashion.

We consider that a cryptocurrency adheres to the efficient market hypothesis when the values of $H_t$ and $C_t$ within a time-window cannot be distinguished from those obtained by chance. To determine the previous condition, we calculate the 95\% random confidence interval (shaded gray in Figs.~\ref{fig:1}B and C) by shuffling the data in each time-window and calculating $H$ and $C$ for several independent realizations (Methods Section for details). Finally, we define the overall informational efficiency $E$ of a cryptocurrency as the fraction of time at which $H_t$ and $C_t$ are simultaneously within the 95\% random confidence band. The values of $E$ thus range from zero to one, with the lower bound indicating a very low-efficiency cryptocurrency, while the upper bound represents a high-efficiency cryptocurrency. It is worth noting that this definition maps well the ideas underlying the efficient market hypothesis, in the sense that the price of a crypto-asset having a high value of $E$ is expected to be very robust against profitable trading strategies, while the price of a cryptocurrency having a low value of $E$ is more likely to be predicted and vulnerable to betting strategies. For the Bitcoin market, we find $E\approx 0.85$, indicating that this crypto-asset is remarkably adherent to the efficient market hypothesis, much in line with preceding research~\cite{nadarajah2017inefficiency, tiwari2018informational}. This in turn also validates our approach and invites a large-scale analysis along the same lines.

\begin{figure}[!ht]
\centering
\includegraphics[width=0.97\columnwidth, keepaspectratio]{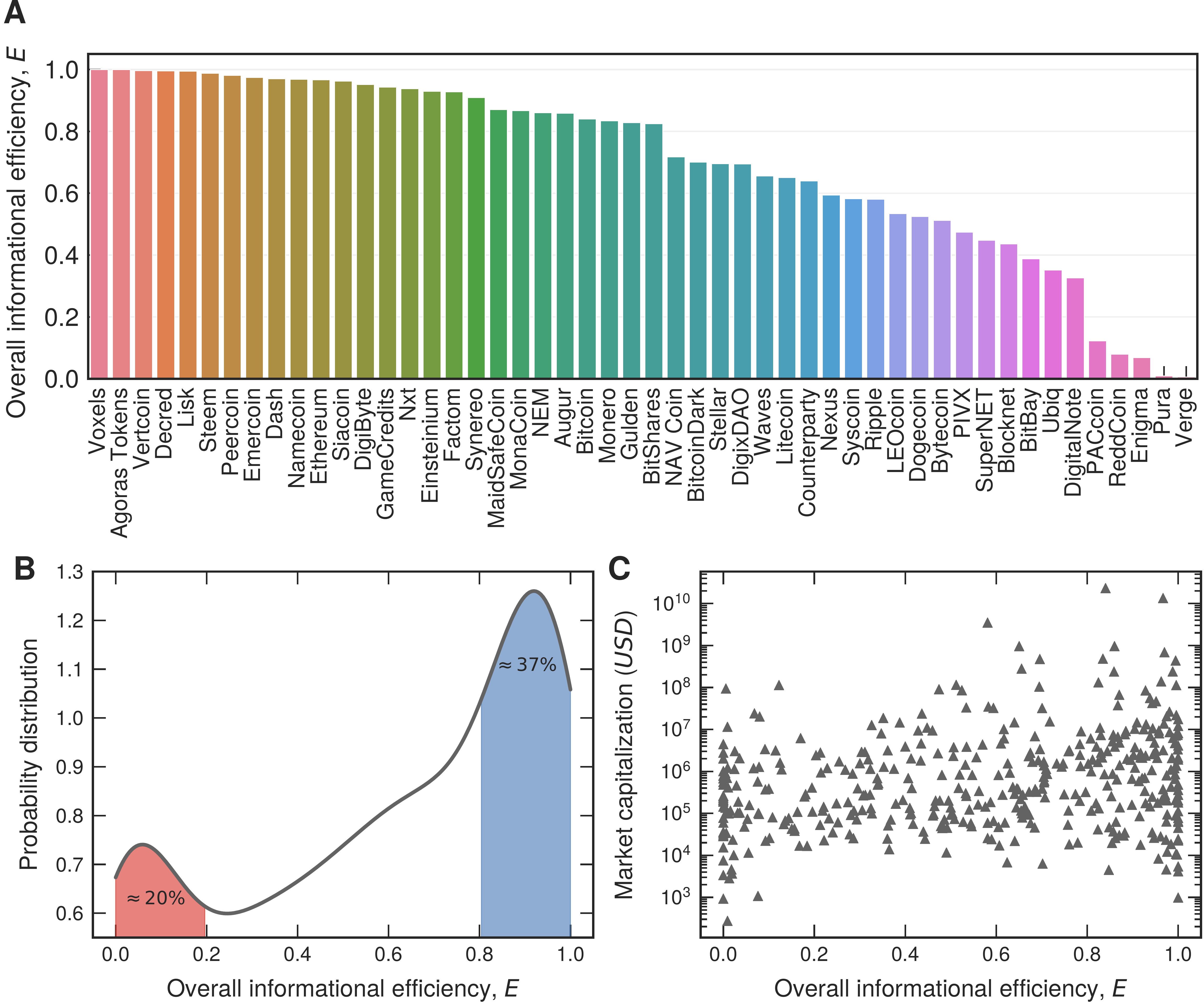}
\caption{The overall efficiency of the cryptocurrency market and its detachment from the market capitalization mean. (A) Overall efficiency $E$ for the top-50 largest cryptocurrencies according to the market capitalization mean. (B) Kernel density estimation of the probability distribution function of the informational efficiency $E$. We note that this distribution is bimodal. The first peak (for $E$ smaller than $0.2$) contains $\approx 20\%$ of the cryptocurrencies (red shaded area), indicating the most informationally inefficient cases, whereas in the second peak (for $E$ greater than $0.8$) we have $\approx 37\%$ of the cryptocurrencies (blue shaded area), the most efficient cases. (C) Scatter plot depicting the values of informational efficiency $E$ versus the market capitalization mean in a linear-log representation. We observe no correlation between these variables, indicating a detachment of informational efficiency from the market capitalization mean. The Pearson linear correlation coefficient is $\approx 0.07$ and the 2-tailed \textit{p}-value is $\approx 0.16$, indicating that the null hypothesis of no linear correlation cannot be rejected. For this analysis, we have considered all the 437 cryptocurrencies with more than 600 observations of $R_t$.}
\label{fig:2}
\end{figure}

To that effect, we calculate the overall efficiency $E$ for or all the 437 cryptocurrencies in our dataset. Figure~\ref{fig:2}A shows this quantity for the top-50 largest cryptocurrencies according to the market capitalization mean, while Fig.~\ref{fig:2}B shows the probability distribution of $E$, as obtained for all the 437 cryptocurrencies in our dataset. It can be observed that this distribution features a strongly pronounced bimodal shape such that the first peak contains $\approx 20\%$ of the cryptocurrencies, while the second peak contains $\approx 37\%$. The first peak corresponds to the most informationally inefficient cryptocurrencies, while the second peak corresponds to the most informationally efficient cryptocurrencies. Evidently, there are significantly more informationally efficient than inefficient digital currencies on the market. Nevertheless, a hard case for a pervading adherence to the efficient market hypothesis would still be quite far-fetched.

To further corroborate our conclusions thus far, we show in Fig.~\ref{fig:2}C how the mean of the market capitalization depends on the informational efficiency $E$. The first thing to notice is that there is hardly any correlation between market capitalization mean and informational efficiency. Indeed, a statistical analysis reveals that the Pearson linear correlation coefficient is $\approx 0.07$, and the 2-tailed \textit{p}-value is $\approx 0.16$. Accordingly, it is impossible to reject the null hypothesis of no linear correlation. Since the market capitalization determines the market value of the cryptocurrency, i.e., the number of crypto-coins multiplied by their current price on the market, we thus find that the adherence of the cryptocurrency market to the efficient market hypothesis is hardly dependent on the volatile price variations in recent years.

\begin{figure*}[!ht]
\centering
\includegraphics[width=0.98\textwidth, keepaspectratio]{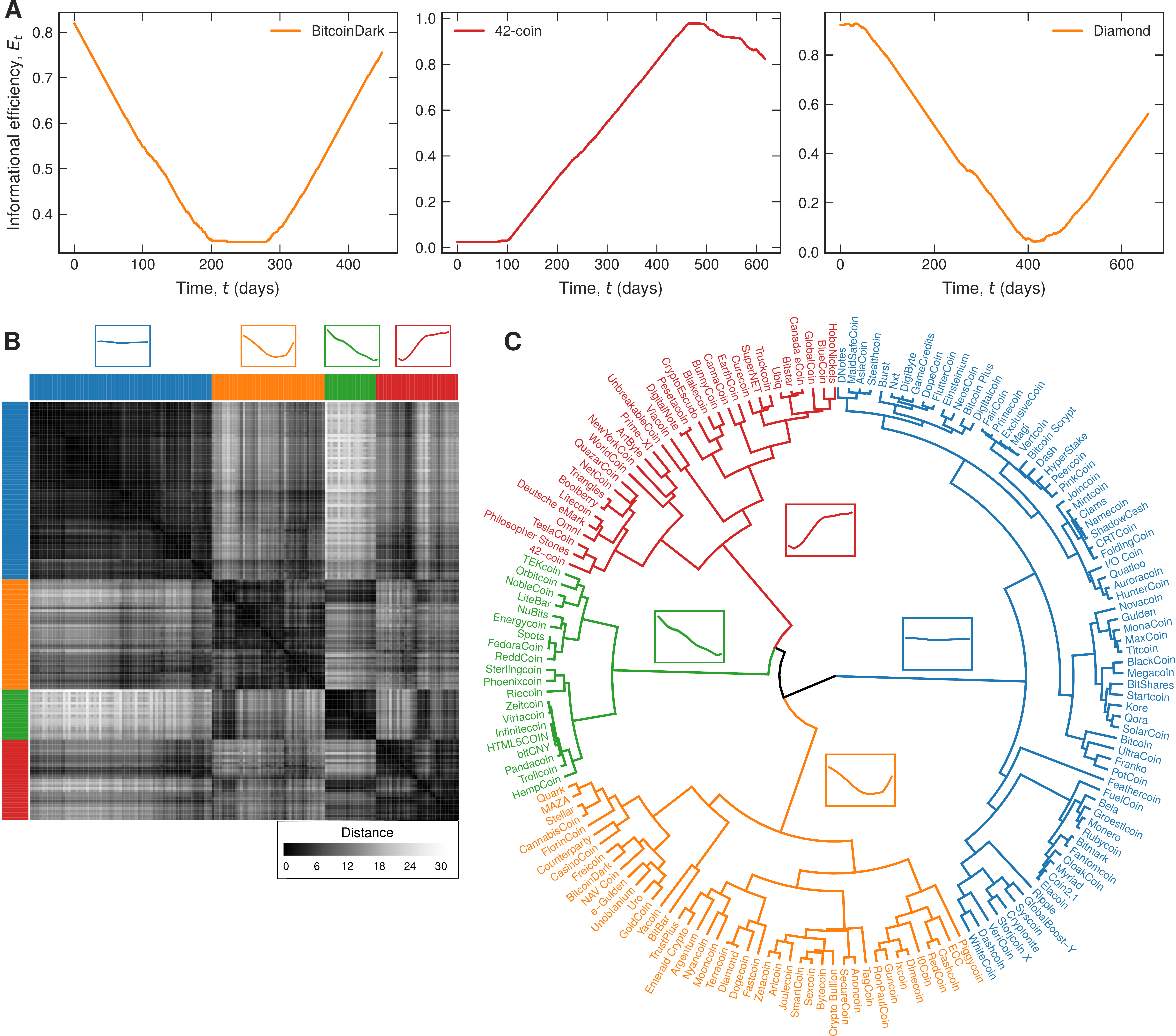}
\caption{Clustering patterns in the dynamics of informational efficiency. (A) Three examples of the time evolution of informational efficiency $E_t$. The dynamics of $E_t$ is obtained by using a 360-day sliding window over the time series $H_t$ and $C_t$ and by calculating the fraction of points (days) that both complexity measures are within the 95\% confidence interval. We note that the motifs of $E_t$ look very similar for BitcoinDark (BTCD) and Diamond (DMD), while 42-coin (42) displays a different trajectory for $E_t$. (B) Matrix plot of the dynamic time warping (DTW) distance among all pairs of the 167 cryptocurrencies having more than 460 observations for $H_t$ and $C_t$. (C) Dendrogram showing the result of the hierarchical clustering based on the DTW distances and using the average linkage criteria. The colored branches indicate the four groups of cryptocurrencies exhibiting similar motifs for the dynamics of $E_t$. These groups are obtained by cutting the dendrogram at the threshold distance that maximizes the silhouette coefficient (Methods Section for details). The order of rows and columns in the matrix plot (B) is the same used in the dendrogram, and the colored insets of that panel represent the average pattern of $E_t$ for each group.}
\label{fig:3}
\end{figure*}

We now focus on a fine-grained view of the dynamical behavior of the informational efficiency. To do so, we select the 167 cryptocurrencies having more than 460 observations for $H_t$ and $C_t$. From these time series, we define the time-dependent informational efficiency $E_t$ by using a 360 data points sliding window which moves forward one day at a time and calculating the fraction of days at which $H_t$ and $C_t$ are simultaneously within the 95\% random confidence interval. These new time series $E_t$ thus provide a detailed view about how the informational efficiency of cryptocurrencies changes over time, which in turn gives us a chance to find those currencies with similar temporal characteristics. Figure~\ref{fig:3}A shows three examples of the time evolution of informational efficiency $E_t$ for BitcoinDark (BTCD), 42-coin (42) and Diamond (DMD). This small sample already indicates that the shape of $E_t$ can be very similar among some cryptocurrencies such as for BitcoinDark and Diamond. To extend this comparative analysis to all 167 cryptocurrencies, we use the dynamic time warping algorithm~\cite{sakoe1978dynamic} (DTW, Methods Section for details) for measuring similarity among all possible pairs of $E_t$ time series. To state briefly, DTW is a distance-like and shape-based similarity measure between two time series that can be used to compare time series of different lengths. For the examples of Fig.~\ref{fig:3}A, the DTW distance between BitcoinDark and Diamond is $3.64$, while the distance between 42-coin and these two cryptocurrencies are $7.73$ and $11.72$ respectively. Thus, the closer the DTW distance between a pair of cryptocurrencies, the more similar is the profile of the evolution of $E_t$ between them. Figure~\ref{fig:3}B shows the matrix plot of the DTW distance between every pair of the selected 167 cryptocurrencies.

To probe for a possible hierarchical organization of cryptocurrencies regarding their informational efficiency evolution, we use the average linkage criteria to build up a dendrogram representation of the DTW distance matrix. This clustering procedure iteratively merges pairs of clusters having the smallest average distance, in which the latter is defined as the average value of the distance between all pairs of elements among the two clusters. Figure~\ref{fig:3}C depicts this dendrogram that supports the idea that cryptocurrencies are hierarchically organized regarding the profile of the evolution of their informational efficiency $E_t$. This hierarchical organization also gives us a chance to find groups of currencies with similar temporal characteristics. To do so, we need to find an optimal threshold distance for cutting the dendrogram and splitting the cryptocurrencies into groups. A `natural' approach to specify this threshold distance is by maximizing the silhouette coefficient, a clustering evaluation metric that simultaneously grades the cohesion and the separation of the produced groups (Methods Section for details). By maximizing the silhouette coefficient, we find that $5.7$ is the optimal DTW distance that splits the dendrogram into four groups of cryptocurrencies indicated by the colored branches in Figure~\ref{fig:3}C. Thus, cryptocurrencies belonging to the same group have roughly the same temporal profile of the informational efficiency $E_t$, whose average behavior corresponds to the four insets at the top of Fig.~\ref{fig:3}B. These four average profiles can be qualitatively described by: $i)$ an almost constant and high-efficiency level (blue group, 43\% of cryptocurrencies); $ii)$ an informational efficiency that starts at a higher level, decreases to a lower one, and increases again (orange group, 26\% of cryptocurrencies); $iii)$ a decreasing informational efficiency (green group, 12\% of cryptocurrencies); and $iv)$ an increasing informational efficiency (red group, 19\% of cryptocurrencies).

\begin{figure}[ht]
\centering
\includegraphics[width=0.68\textwidth, keepaspectratio]{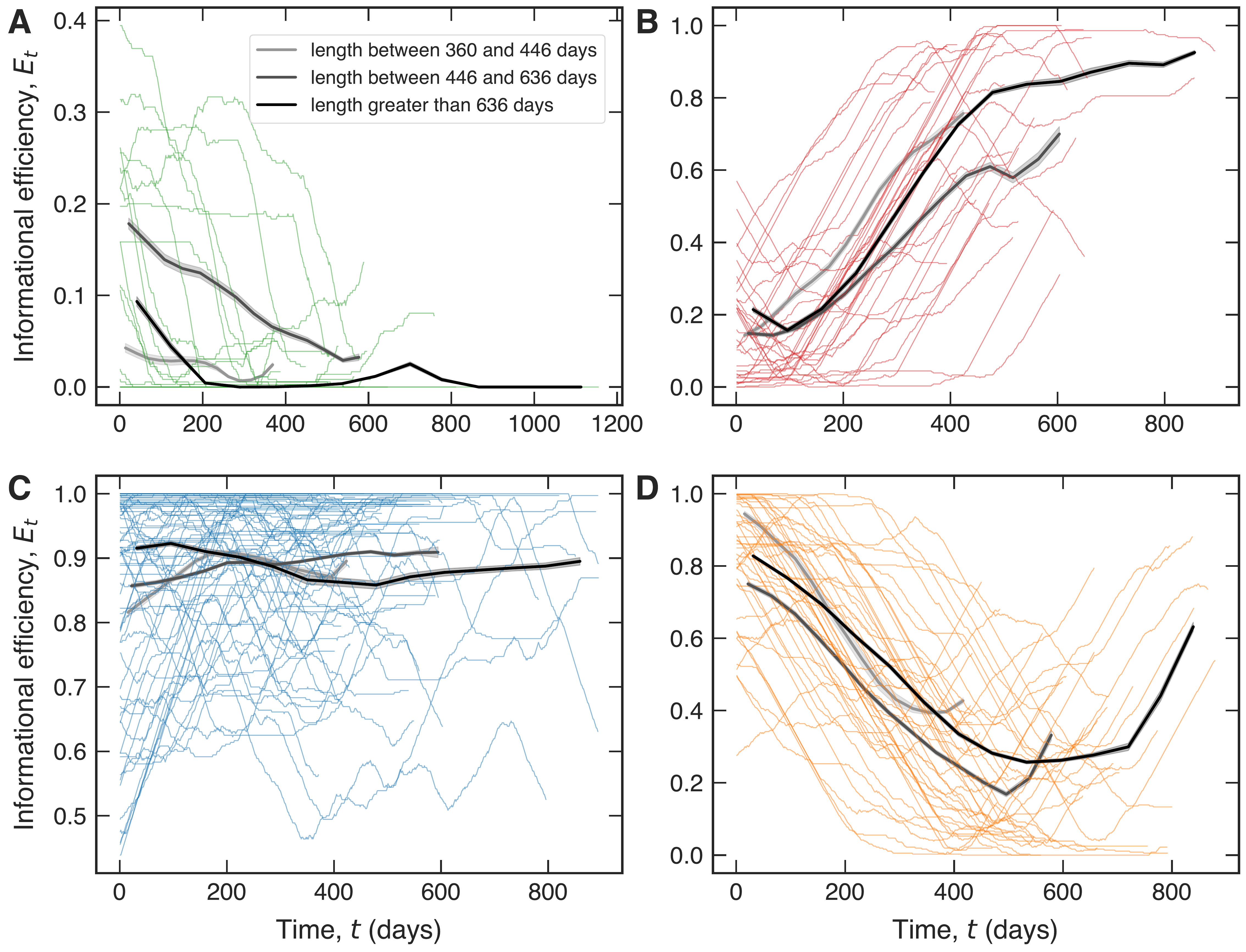}
\caption{Temporal evolution of informational efficiency within the four different groups of cryptocurrencies. Panels (A) to (D) show the time series of the informational efficiency $E_t$ for the four groups of cryptocurrencies obtained by the hierarchical clustering procedure. The colored curves represent each one of the 167 cryptocurrencies, while the gray curves show the average behavior of $E_t$ in each group for three intervals of lengths of these series. These length intervals were chosen to contain approximately the same number of cryptocurrencies. In (A) we observe predominantly a decreasing pattern for $E_t$, while in (B) we have the prevalence of an increasing trend for the youngest cryptocurrencies in that group until it starts saturating, in particular for the older cryptocurrencies. In (C) there is a varying behavior of the informational efficiency $E_t$ over time, that on average is nearly constant and highly efficient. Finally, in (D) we observe that in the beginning most cryptocurrencies are highly efficient, but then follow a decreasing trend before they recover towards the end of the observation time. These results suggest that the younger cryptocurrencies in each group are likely to follow the trend of their elders.}
\label{fig:4}
\end{figure}

Lastly, it is revealing to look at the temporal evolution of informational efficiency $E_t$ for each of the four different clusters, such that the currencies with different ages are distinguished, as shown in Fig.~\ref{fig:4}. We observe in all four groups that the trends of older currencies appear to be followed up by mid-age and younger currencies. Accordingly, this dynamic analysis of informational efficiency over time reveals that younger currencies in each group are poised to follow the trend of their `elders'. It is also worth noting that 81\% of the cryptocurrencies belong to the groups characterized by a constant and high informational efficiency or by an increasing efficiency level (blue, red, and orange groups). In addition to that, among the top-10 largest currencies according to the market capitalization mean, we verify that six belong to the blue group [BitShares (BTS), Bitcoin (BTC), Dash (DASH), MaidSafeCoin (MAID), Monero (XMR), Ripple (XRP)], three to the orange group [Bytecoin (BTS), Dogecoin (DOGE), Stellar (XLM)] and one to the red group [Litecoin (LTC)]. This in turn suggests that the younger cryptocurrencies, which currently do not abide to the efficient market hypothesis, might very well do so in the foreseeable future, and at that time render the digital currency market equally compliant with the efficient market hypothesis as the traditional financial markets usually are. The relatively small deviations of younger currencies in comparison to the informational efficiency trends of the older currencies further suggest that this coming-of-age of the digital currency market might come rather soon.

\section*{Discussion}
We have presented a large-scale efficiency analysis of the cryptocurrency market, focusing on temporal clustering patterns and the coming-of-age of the digital currency market in general. By using permutation entropy and statistical complexity over sliding time-windows of price log returns, our research reveals that the cryptocurrency market is to a quite significant degree compliant with the efficient market hypothesis, with only 20\% of cryptocurrencies being less than 20\% informationally efficient. On the other hand, over half of the cryptocurrency market is over 60\% informationally efficient, and 37\% of the 437 cryptocurrencies in our study is over 80\% informationally efficient. This is in accord with preceding research on the same subject~\cite{nadarajah2017inefficiency, tiwari2018informational}, where authors have reported similar conclusions for the Bitcoin currency.

Moreover, based on a dynamic analysis of informational efficiency over time, and by clustering together cryptocurrencies with similar temporal profiles, we have revealed which factions of the digital currency market follow the same ups and downs with respect to their informational efficiency. While the particularities of the makeup of different clusters might be of interest to investors seeking a diversified portfolio, we have shown that within the four identified clusters the younger currencies appear to follow the trend of their `elders'. We have argued that this can be interpreted as the coming-of-age of digital currencies, such that younger cryptocurrencies, which as of yet do not abide to the efficient market hypothesis, might do so in the future. The similarities in temporal trends between younger and older cryptocurrencies led us to conclude that the coming-of-age in this regard is certainly more imminent than far-fetched.

Taken together, we hope that our physics-inspired approach aimed at determining the efficiency of the cryptocurrency market in its current state will continue to inspire further fruitful synergies between physics and economics~\cite{mantegna1999introduction}, and we hope that the presented results will contribute to the lively research environment of the digital currency market.

\section*{Methods}
\subsection*{Data}\label{data}
We have obtained the daily closing prices and the market capitalization of 1509 cryptocurrencies by crawling the website \url{coinmarketcap.com} on 13 February 2018. These data span different periods ranging from a few days (e.g., for the cryptocurrencies Medicalchain and Farstcoin) to almost five years (for Bitcoin). We have selected the 437 cryptocurrencies having more than 600 observations for the analysis of the overall informational efficiency, while the results related to the dynamical behavior of the informational efficiency are based on 167 cryptocurrencies spanning more than 960 days. These filters are necessary to have a reliable estimate of the informational efficiency in each analysis, ensuring that the estimation of the overall efficiency is based on at least 100 observations of entropy and complexity, and that the time series of the informational efficiency are longer than 100 days.

\subsection*{Permutation entropy and statistical complexity}\label{permutation}
The permutation entropy~\cite{BandtPompe2002} and the permutation statistical complexity~\cite{LopezManciniCalbet1995,RossoLarrondoMartinPlastinoFuentes2007} are complexity measures originally proposed for characterizing time series. Both quantities are based on a probability distribution related to the local ordering patterns among consecutive time series elements. To define these measures, let us consider a time series $\{x_t\}_{t=1,2,\ldots,n}$ and overlapping partitions of length $d>1$ (the embedding dimension) represented by
\begin{equation}
(\vec s\,)\mapsto (x_{s-(d-1)},x_{s-(d-2)},\ldots,x_{s-1},x_s),
\end{equation}
where $s=d,d+1,\ldots,n.$ For each one of these $(n-d+1)$ partitions, we investigate the $d!$ permutations $\pi = (r_0,r_1,\ldots,r_{d-1})$ of $(0,1,\ldots,d-1)$ defined by the ordering $x_{s-r_{d-1}}\leq x_{s-r_{d-2}}\leq \ldots \leq x_{s-r_0}$. These permutations represent all $d!$ possible ordering patterns among the $d$ elements of the $(\vec s\,)$ partitions. We thus calculate the relative frequency of each one of these $d!$ permutations
\begin{equation}
p(\pi_i) = \frac{\mbox{the  number of $s$ that has type } \pi_i}{(n-d+1)},
\end{equation}
defining the probability distribution of the ordinal patterns $P=\{p(\pi_i)\}_{i=1,2,\ldots,d!}$.

The permutation entropy is thus defined as a normalized Shannon entropy~\cite{Shannon1948} of $P$, that is,
\begin{equation}
H(P)=-\frac{1}{\ln(d!)}\sum_{i=1}^{d!}p(\pi_i)\ln p(\pi_i),
\end{equation}
where the $\ln(d!)$ is a normalization constant (the maximum value of the Shannon entropy). The values of $H$ quantify the degree of disorder in the ordering dynamics of the time series elements. Values of $H\approx1$ indicate that elements of $x_t$ are locally randomly ordered, while values of $H\approx0$ show that these elements appear almost always in a particular order.

The permutation statistical complexity is defined as
\begin{equation}
C(P)=\frac{D(P,U)H(P)}{D^*},
\end{equation}
where $D(P,U)$ is a relative entropic measure (the Jensen-Shannon divergence) between $P=\{p(\pi_i)\}_{i=1,2,\ldots,d!}$ and the uniform distribution $U = \{u_i=1/d!\}_{i=1,2,\ldots,d!}$, that is,
\begin{equation}
D(P,U) = S\left(\frac{P+U}{2}\right) - \frac{S(P)}{2} - \frac{S(U)}{2},
\end{equation}
and $D^*$ is a normalization constant obtained by calculating $D(P,U)$ when $P=\{p_i=\delta_{1,i};~i=1,\ldots,d!\}$. The statistical complexity quantifies the degree of structural complexity present in the time series. A value of $C\approx0$ occurs for both extremes of order and disorder, whereas $C>0$ represents more complex patterns in the arrangement of the series elements.

We note that both measures depend only on the choice of the embedding dimension $d$. However, this choice is not entirely arbitrary because the condition $d!\ll n$ must be satisfied to obtain reliable statistics. Because of their simplicity, intuitive meaning, and scalability from the computational point of view, this framework has been successfully used in several applications with time series~\cite{ribeiro2012complexity,zunino2012review,PhysRevE.89.012905,Jovanovic2016,Stosic20161136,ribeiro2017characterizing} and image analysis~\cite{schlemmer2015quantifying,antonelli2017permutation,sigaki_pnas18,antonelli2018mammographic}. 

\subsection*{Measuring information efficiency}\label{informational}
The information efficiency of each cryptocurrency is estimated from the logarithmic daily price returns series defined by~\cite{mantegna1999introduction}
\begin{equation}
R_t = \log P_t-\log P_{t-1},
\end{equation}
where $\log P_t$ and $\log P_{t-1}$ are the natural logs of closing prices at time $t$ and $t-1$.

We calculate the values of $H$ and $C$ within a sliding window of size $w$ moving forward in daily steps over the return series. This procedure defines new time series representing the values of $H_t$ and $C_t$ in each window, where $t$ stands for center of time-windows (Figs.~\ref{fig:1}B and~\ref{fig:1}C). We have used $w=500$ days to satisfy the condition $d!\ll n$. We have also estimated the 95\% random confidence intervals by shuffling the time series in each window and calculating the values of $H$ and $C$ for 30 independent realizations. We define the overall information efficiency $E$ of a given cryptocurrency as the fraction of time that the values of entropy and complexity stay simultaneously within the random confidence band. The information efficiency time series $E_t$ is estimated by calculating the same fraction within a sliding window of 360 days, as illustrated in Fig.~\ref{fig:3}A. We also observe that this procedure is robust against different sizes for the sliding window, for instance, we have verified that very similar results are obtained for windows with 30 and 180 days.

\subsection*{Hierarchical clustering procedure}\label{cluster}
We have used the dynamic time warping (DTW)~\cite{sakoe1971dynamic} algorithm for estimating the similarities between the dynamical behavior of $E_t$ among the cryptocurrencies. DTW is a distance-like and shape-based similarity measure that can be applied to time series of different lengths and range of values. This method calculates an optimal alignment between two time series by minimizing a cost function (or distance)~\cite{sakoe1978dynamic} and it is also widely used for time-series clustering~\cite{aghabozorgi2015time}. We construct a matrix of the DTW distances between every pair of cryptocurrencies (Fig.~\ref{fig:3}B) and use the average linkage criteria~\cite{hastie2013elements} to hierarchically cluster the cryptocurrencies, as shown in Fig.~\ref{fig:3}C. This clustering procedure iteratively merges clusters having the smallest average distance. The threshold distance used to determine the number of clusters was obtained by maximizing the silhouette coefficient~\cite{rousseeuw1987silhouettes}. This coefficient quantifies the consistency of the clustering procedure and is defined by the average value of
\begin{equation}
s_i = \frac{b_i-a_i}{\max(a_i,b_i)}\,,
\end{equation}
where $a_i$ is the cohesion (the average intracluster distance) and $b_i$ is the separation (the average nearest-cluster distance) for the $i$-th cryptocurrency. The higher the average value of the silhouette for all cryptocurrencies, the better the cluster configuration. We find that the maximum value of the silhouette coefficient occurs for the threshold distance $5.7$, which is the distance used to obtain the four clusters depicted in Figs.~\ref{fig:3}B and C. The algorithms used for evaluating the DTW and performing the hierarchical clustering procedure are implemented in the \textit{Python} modules \textit{dtaidistance}~\cite{dtwdistance} and \textit{SciPy} library~\cite{scipy} respectively.


\section*{Acknowledgements}

This research was supported by Conselho Nacional de Desenvolvimento Cient\'ifico e Tecnol\'ogico (CNPq), Coordena\c{c}\~ao de Aperfei\c{c}oamento de Pessoal de N\'ivel Superior (CAPES) Grants 440650/2014-3 and 303642/2014-9, and by the Slovenian Research Agency Grants J1-7009, J4-9302, and J1-9112.

\section*{Author contributions statement}

HYDS, MP and HVR designed research, performed research, analyzed data, and wrote the paper.

\section*{Additional information}

\textbf{Competing interests:} The authors declare that they have no conflict of interest.

\end{document}